\documentclass[12pt]{article}
\usepackage{jpc}
\usepackage{amsmath}
\usepackage{epsfig}
\usepackage{cite}
\makeatletter
\def\@cite#1#2{$^{\mbox{\scriptsize #1\if@tempswa , #2\fi}}$}
\makeatother
\newdimen\titlewidth
\titlewidth \textwidth
\advance\titlewidth-25mm

\newcommand   {\Cv}    {C_V}
\newcommand   {\Rg}    {R_\text{g}}
\newcommand   {\na}    {n_\alpha}
\newcommand   {\nb}    {n_\beta}
\newcommand   {\chia}  {\chi_\alpha}
\newcommand   {\chib}  {\chi_\beta}
\newcommand   {\la}     {\lambda_\alpha}
\newcommand   {\lb}     {\lambda_\beta}
\newcommand   {\kB}    {k_\text{B}}
\newcommand   {\ev}[1] {\langle #1\rangle}
\begin{document}
\begin{center}
\begin{center}
\noindent{\Large\textbf{Differences in Solution Behavior among\\[2mm] 
Four Semiconductor-Binding Peptides}}\\[10mm]   
\begin{large}
Simon Mitternacht,\footnote[2]{Lund University}
Stefan Schnabel,\footnote[3]{University of Leipzig} 
Michael Bachmann,$^{\dag,\ddag}$\\ 
Wolfhard Janke,$^{\ddag}$ and 
Anders Irb\"ack\footnote[1]{Corresponding author. E-mail: anders@thep.lu.se; tel.: +46 46 2223493;
fax: +46 46 2229686}$^{,\dag}$\\[10mm]
\end{large}
{\em Computational Biology \& Biological Physics Group, 
Department of Theoretical Physics, Lund University,  
S\"olvegatan 14A, SE-223 62 Lund, Sweden, and\\
Institute for Theoretical Physics and Centre for Theoretical Sciences (NTZ), 
University of Leipzig, Augustusplatz 10/11, D-04109 Leipzig, Germany}\\[10mm]
\end{center}
\end{center}
\newpage

\begin{center}
{\bf Abstract}
\end{center}
Recent experiments have identified peptides with adhesion affinity
for GaAs and Si surfaces. Here we use all-atom Monte Carlo (MC) simulations 
with implicit solvent to
investigate the behavior in aqueous solution of four such peptides, all with 
12 residues. At room temperature, we find that all the four peptides 
are largely unstructured, which is consistent with experimental data. 
At the same time, we find that one of the peptides is structurally different
and more flexible, compared to the others. This finding points at 
structural differences as a possible explanation for differences in adhesion 
properties between these peptides. By also analyzing designed mutants
of two of the peptides, an experimental test of this hypothesis is proposed.

\vspace{24pt}


\newpage 

%
\section*{\centering{Introduction}}
The advancing progress in manipulating proteins and non-biological 
macromolecules and materials at the nanometer scale opens up possibilities 
for constructing novel hybrid materials with potential applications in
bionanotechnology.~\cite{sarikaya1,gray} An important
development in this direction is the identification of proteins that can 
bind to specific compounds. Over the last decade, genetic 
engineering techniques have been successfully employed to find peptides with 
affinity for, e.g., metals,\cite{brown1,sano} semiconductors\cite{whaley1} 
and carbon nanotubes.\cite{wang} However, the mechanisms by which peptides
bind to these materials are not completely understood; e.g., it is unclear 
what role conformational changes play in the binding process.     

Here we report atomic-level simulations of the solution behavior 
of four 12-residue peptides, whose adhesion properties to  
(100) surfaces of GaAs and Si crystals were studied in recent 
experiments.\cite{whaley1,goede1,goede2} The main quantity measured  
in the experiments was the peptide adhesion coefficient (PAC), defined as 
the percentage of surface coverage, after drying and washing of 
the samples which were originally in contact with the peptide solution. 
This quantity was measured by AFM for the different peptide-substrate 
combinations,\cite{goede1,goede2} and was found to show a clear 
dependence on both peptide and substrate (see below).  

How the binding occurs in these peptide-surface systems 
is unclear. However, although the bound peptides were 
found to form clusters,\cite{goede2} it seems unlikely that the peptides
aggregate before binding to the surface, because the hydrophobicity of the 
peptides studied
is low and the peptide concentration was low, in the nanomolar range. 
A more accurate description is probably that the 
peptides bind one by one, a process that, in principle, can occur in two 
fundamentally different ways. One possibility is a docking 
behavior, where the peptides bind to the surface without undergoing 
any major conformational change. This scenario assumes that the
peptides have a stable structure in solution, and that this 
structure matches the structure of the surface, e.g., with respect to 
polarization. The second variant is that the peptide is unstructured 
before binding occurs. Although the bound peptide structure need not be 
unique, the process would then have similarities with coupled 
folding-binding,\cite{dyson} which can be an efficient mode 
of binding compared to docking.\cite{shoemaker,irbaeck1}
Measurements of circular dichroism (CD) spectra suggest that all the 
four studied peptides are largely unstructured in solution,\cite{goede2} 
thus favoring the second type of binding over docking. 

Recent studies have found that the adhesion propensity
of peptides to various surfaces can be in part explained in terms
of adhesion properties of their constituent amino 
acids,\cite{willett1,peelle1} However, the amino acid composition 
alone cannot explain the PAC values obtained experimentally for the 
four peptides studied here. In fact, two of these peptides share
exactly the same amino acid composition, but still have quite different 
adhesion properties. In order to explain the adhesion properties,
it might thus be necessary to take structural characteristics into 
account. However, as already indicated, the CD measurements did not  
reveal any clear structural differences between these 
peptides.\cite{goede2}

The aim of our study is to get a more detailed picture of the behavior 
in aqueous solution of these peptides, and look for possible 
structural differences not seen in the CD analysis. A perfect 
model for folding simulations does not exist. It is worth
noting, however, that the model we use,\cite{irbaeck2,irbaeck4} 
despite a simplified energy function, 
is capable of folding both $\alpha$-helical and $\beta$-sheet 
peptides, without changing any model parameters.\cite{irbaeck4} 

Three of the peptides we study have previously been simulated\cite{gbcj1}   
using the ECEPP/3 force field.\cite{ecepp} This study found only minor
differences in folding behavior between these peptides. To further elucidate
the structural properties of these peptides at room temperature, we here 
perform simulations using an alternative model, which has given realistic  
results for the stability and its temperature dependence for the peptides
that it was able to fold.\cite{irbaeck4}

Simulating the actual binding of the peptides to the surface 
is more challenging due to uncertainties about the precise form of 
the peptide-surface interactions and their dependence on
solvation effects.\cite{cormack,liang} Nevertheless, such simulations 
have been performed for gold-binding peptides.\cite{schulten1}
The phase structure for chain adsorption to attractive surfaces
has been investigated using lattice models for 
polymers\cite{vrbova1,bj1,bj2} and peptides.\cite{bj3} Simplified
statistical-mechanical models have also been 
used to study molecular recognition of
patterned 
surfaces,\cite{muthukumar,bratko,golumbfskie,bogner1,kriksin,moghaddam} 
and conformational changes of proteins adsorbed to a 
solid surface.\cite{zhdanov,castells} 

\section*{\centering{Model and Methods}}
\label{sec:mod}
\paragraph{Peptides Studied} 
The four peptides we study are listed in Table~1, 
where also PAC values for (100) GaAs and Si surfaces 
can be found. The sequence S1 was selected from a huge 
library of 12-mers for adhesion to GaAs.\cite{whaley1}
Its very poor propensity to adhere to Si is noteworthy. The 
sequence S2 is obtained from S1 by exchanging two histidines for 
alanines. This double mutation leads to a reduced PAC for 
GaAs  and a slightly increased PAC for Si. The peptide
adhering best to the Si surface is S3,  which is a random 
permutation of S1. The Si PAC is a factor 15 higher for S3 
than for S1, despite that their amino acid composition
is the same. The last sequence, S4, is derived from S1 by 
replacing three asparagines by alanines. This change
results in a slightly reduced GaAs PAC and an increased 
Si PAC.

\paragraph{Peptide Model}
The model we use contains all atoms of the peptide chain, including 
H atoms, but no explicit water molecules. It assumes fixed bond 
angles, bond lengths and peptide torsion angles ($180^\circ$), so that
each amino acid has the Ramachandran angles $\phi,\psi$ and a
number of side-chain torsion angles as its degrees of freedom. 
Here a brief presentation of the energy function will be given. 
Detailed descriptions  of the parameterization 
of the geometry\cite{irbaeck2} and the different 
energy terms\cite{irbaeck4} can be found elsewhere. 

The energy function consists of four terms,
\begin{equation}
  E = E_\text{ev} + E_\text{loc} + E_\text{hb} + E_\text{hp}\,.
\end{equation}
The first term, $E_\text{ev}$, represents excluded volume effects 
and is of the form
\begin{equation}
  E_\text{ev} = \kappa_\text{ev} \sum_{i<j} 
  \biggl[\frac{\lambda_{ij}(\sigma_i + \sigma_j)}{r_{ij}}\biggr]^{12}\,,
\end{equation}
where the sum is over all atom pairs. The parameters $\sigma_i$ 
are atomic radii and $\lambda_{ij}$ is a scale factor, which  
is 1.0 for pairs connected by three covalent bonds and 0.75 otherwise.

The second term represents an interaction between neighboring 
NH and CO partial charges along the backbone. It is given by 
\begin{equation}
  E_\text{loc} = \kappa_\text{loc} \sum_I {
    \biggl[\sum_{
	\begin{smallmatrix} 
	  i=\text{N,H}\, \in I \\ 
	  j=\text{C,O}\, \in I 
	\end{smallmatrix}
      } 
      \frac {q_i q_j} {r_{ij}} \biggr]
  }
  \,,
\end{equation}
where the outer sum is over all amino acids and the $q_i$ are partial
charges.

The H bond contribution, $E_\text{hb}$, consists of two parts:
backbone-backbone bonds and backbone-sidechain
bonds,
\begin{equation}
  E_\text{hb} = \epsilon_\text{hb}^{(1)}
  \sum_\text{bb-bb} u(r_{ij})v(\alpha_{ij},\beta_{ij}) +\\
  \epsilon_\text{hb}^{(2)}
  \sum_\text{bb-sc} u(r_{ij})v(\alpha_{ij},\beta_{ij})\,,
\end{equation}
where $r_{ij}$ denotes the HO distance, $\alpha_{ij}$ the NHO angle and
$\beta_{ij}$ the HOC angle. The function $u(r)$ is given by
\begin{equation}
  u(r) = 5\,\biggl(\frac{\sigma_\text{hb}}{r}\biggr)^{12} -
  6\, \biggl(\frac{\sigma_\text{hb}}{r}\biggr)^{10}
\end{equation}
and the angular dependence is
\begin{equation}
  v(\alpha,\beta) = \begin{cases} (\cos \alpha \cos \beta)^{1/2} &
    \text{if }\alpha,\beta>90^\circ, \\ 0 & \text{otherwise.}
      \end{cases}
\end{equation}

The last energy term, $E_\text{hp}$, represents an effective hydrophobic 
attraction and has the form
\begin{equation}
  E_\text{hp} = - \sum_{I<J} M_{IJ}C_{IJ}\,,
\end{equation}
where the sum is over all pairs of nonpolar amino acids. 
The $M_{IJ}$ ($\ge0$) are constants that determine the strength of 
attraction between amino acids $I$ and $J$. 
$C_{IJ}$ is a geometric factor and a measure of the degree 
of contact between two side chains. It is defined as
\begin{equation}
    C_{IJ} = \frac{1}{N_I + N_J} \Biggl[\, \sum_{i\in A_I} f(\min_{j\in A_J} r_{ij}^2)
    + \sum_{j\in A_J} f(\min_{i\in A_I} r_{ij}^2) \Biggr]\,,
\end{equation}
where $A_I$ denotes a predefined set of $N_I$ sidechain atoms for residue $I$. 
The function $f(x)$ is given by $f(x)=1$ if $x<A$, $f(x)=0$ if $x>B$,
and $f(x)=(B-x)/(B-A)$ if $A<x<B$ [$A=(3.5\,{\rm \AA}){}^2$ and
$B=(4.5\,{\rm \AA}){}^2$].

\paragraph{Simulation Method}
To investigate the solution behavior of the peptides S1--S4, we perform    
simulated-tempering\cite{Marinari:92,Lyubartsev:92} simulations with eight 
temperatures in the range 275--369\,K, and some reference runs at a 
constant temperature of 1\,000\,K. The conformational updates we use are 
rotations of single backbone and sidechain torsion angles, and a 
semi-local backbone update, biased Gaussian steps (BGS),\cite{BGS} 
which updates seven or eight consecutive angles in a manner 
that keeps the rest of the molecule approximately fixed.
In the simulated-tempering runs these updates are called in different 
proportions at different temperatures with more BGS at lower 
temperatures. At 299\,K, the fractions of attempted single-angle backbone moves,
sidechain moves and BGS are 0.29, 0.51 and 0.20, respectively. 
In the 1\,000\,K simulations the corresponding 
fractions are 0.245, 0.51 and 0.245.
 
Our simulations are carried out 
using the software package PROFASI,\cite{irbaeck5}
which is a C++ implementation of the above model. 
Each simulation comprises $10^9$ elementary MC steps. 

The results of our simulations are analyzed using multi-histogram
techniques.\cite{ferrenberg} All statistical uncertainties quoted are 
1$\sigma$ errors obtained by the jackknife method.\cite{miller} 
\section*{\centering{Results and Discussion}}
\label{sec:dis}
\paragraph{Overall Structure and Temperature Dependence}
Cooperative structural activity is typically signaled by a peak in the 
statistical fluctuations of system relevant quantities, such as the 
energy. Figure~1 shows how the  
specific heat, $\Cv=d\ev{E}/dT=(\ev{E^2}-\ev{E}^2)/\kB T^2$, 
and the temperature derivative of the radius of gyration, $d\ev{\Rg}/dT$, 
vary with temperature for the sequences S1--S4 ($\ev{\cdot}$ denotes a 
Boltzmann average). The qualitative behavior of the three sequences 
S1, S2 and S4 is virtually identical. 
For all three sequences, the specific heat exhibits a broad peak 
with maximum around 280\,K. The $d\ev{\Rg}/dT$ curves show a similar
broad peak, although the statistical errors are larger and the 
maximum is slightly shifted toward higher temperature.

In the temperature regime where these peaks occur, it turns 
out that the secondary-structure content of these three sequences
changes relatively rapidly. As the temperature decreases, the    
$\alpha$-helix content, $\ev{\na}$, increases, whereas the
$\beta$-strand content, $\ev{\nb}$, decreases slightly, as can be seen 
from Fig.~2. 
These results indicate that the structures 
with lowest energy are $\alpha$-helical for S1, S2 and S4. 
It should be noted, however, that the $\alpha$-helix content remains  
small, $<0.25$, all the way down to 273\,K.        

The sequence S3 shows a markedly different behavior. Neither 
$\Cv$ nor $d\ev{\Rg}/dT$ has a maximum within the temperature range studied;
both quantities increase monotonically with decreasing temperature 
(see Fig.~1). Furthermore, the $\beta$-strand content
remains larger than the $\alpha$-helix content at low temperature for this
sequence (see Fig.~2). The $\beta$-strand content does not 
decrease with decreasing temperature, and the $\alpha$-helix content 
increases much less than for the other sequences.     

Figure~3 shows typical low-energy conformations for the
four different sequences, as obtained by simulated 
annealing.\cite{kirkpatrick} 
As one might expect from the temperature dependence of the $\alpha$-helix
and $\beta$-strand
contents, the structure is $\alpha$-helical for S1, S2 and S4.
However, the $\alpha$-helix does not span the entire chain, but  
rather the region between residues 3 to 12. That the beginning of the
sequence does not make $\alpha$-helix structure is not unexpected, because 
there is a proline at position 4. The lowest-energy structure we find for S3 
is a $\beta$-hairpin. Its turn is at residues 6 and 7. The second strand of 
the $\beta$-hairpin, spanning residues 8--12, is not perfect but broken in the
vicinity of the proline at position 9.

It must be stressed that the states illustrated in Fig.~3
are only weakly populated at room temperature, as is evident from the    
secondary-structure contents shown in Fig.~2. Our results
are thus consistent with the CD analysis of the solution behavior
of these peptides,\cite{goede2} at room temperature and pH 7.6,
which suggests that they all are largely unstructured.

Our conclusion that the $\alpha$-helix content, at low temperature, 
is higher than the $\beta$-strand content for S1 and S2, 
is in agreement with a previous study of S1--S3 based on
the ECEPP/3 force field.\cite{gbcj1}   
However, in that study, the
sequence S3 was found to be $\alpha$-helical as well. Furthermore, the 
$\alpha$-helix content of S1 and S2 was significantly higher 
compared to what we find and to what is indicated by   
the CD results.\cite{goede2} 

Having studied the overall structure and the temperature 
dependence, we now turn to a more detailed structural 
description at $T=299$\,K, which is close to where the CD 
measurements were taken.\cite{goede2} This discussion  
will mainly focus on S1 and S3, as the double mutant S2 and 
the triple mutant S4 show a behavior very similar to that 
of S1. 

\paragraph{Structural Characterization at $\boldsymbol{T=299}$\,K} 
To further elucidate the structure and free-energy landscape  
of these peptides, we analyze root-mean-square 
deviations (RMSD) from suitable reference structures (calculated
over backbone atoms). We first consider an $\alpha$-helical 
reference structure. The N-terminal part of S1 is rather flexible 
due to a proline at position 4. 
Similarly, the C-terminal part of S3 is flexible, due to a proline  
at position 9. To reduce noise, we omit these 
tails when calculating RMSD. The reference structure used is an 
$\alpha$-helix with 8 residues. With RMSD calculated this way, we 
study the free energy $F(\Delta,E)$ as a function of RMSD, $\Delta$, and 
energy, $E$, at 299\,K. Figures~4a and 4b 
show contour plots of $F(\Delta,E)$ for S1 and S3. For both 
sequences, the free-energy minimum is at an RMSD of about  
3.4\,\AA, which is approximately the average value for random structures,
as obtained from control runs at 1\,000\,K. This finding supports
the conclusion that S1 and S3 both are largely unstructured at 299\,K. 
A clear local free-energy minimum corresponding to $\alpha$-helix 
structure is 
missing for both sequences. For S1, there is, however, a valley from the 
global minimum in the direction of low RMSD and low energy, and there 
is a small but significant fraction of $\alpha$-helical conformations 
with $\Delta\sim1$\,\AA\ and relatively low energy. For S3, 
there is a valley in the same direction, but it is less 
pronounced, and conformations with a $\Delta$ as small as 1\,\AA\
are rare.  There is also a second valley for S3, where the lowest 
populated energies are found. The appearance of this second valley, 
where $\Delta>3$\,\AA, is not unexpected, given that the lowest-energy 
structure found for S3 is a $\beta$-hairpin (see Fig.~3c).
Figure~4c shows $F(\Delta,E)$ for S3 when 
this $\beta$-hairpin is taken as the reference structure. A local
minimum with $\Delta\sim1$\,\AA\ and low energy can be found, but 
it is very weakly populated. The dominating global minimum corresponds
to unstructured conformations. In fact, the average RMSD from the
$\beta$-hairpin for random S3 conformations, as obtained from a
control run at 1\,000\,K, is about 6\,\AA, which is approximately 
where the global minimum is found at $T=299$\,K.

Next we examine how the $\alpha$-helix and $\beta$-strand 
contents (as defined in the caption of Fig.~2)
vary along the chains. Let $\chia(i)=1$ if residue $i$
is in the $\alpha$-helix state and $\chia(i)=0$ otherwise, so 
that $\ev{\chia(i)}$ is the probability of finding residue $i$      
in the $\alpha$-helix state, and let $\chib(i)$ denote the 
corresponding function for the $\beta$-strand state. Figure~5 
shows $\ev{\chia(i)}$ and $\ev{\chib(i)}$ against $i$ for S1--S4 at 
$T=299$\,K. The low-energy conformations of S1, S2 and S4 shown in    
Fig.~3 contain an $\alpha$-helix starting near position 3 
and ending at the C terminus.
The $\alpha$-helix probability profile  
in Fig.~5a reveals that the stability of this $\alpha$-helix 
is not uniform along the chain; its N-terminal part is most stable, 
whereas the stability decreases significantly toward the C terminus.    
For S3, it can be seen from Fig.~5b that the $\ev{\chib(i)}$
values are similar in the two regions that make the strands of the 
$\beta$-hairpin in Fig.~3c. An exception is Pro9, for which
$\ev{\chib}(i)$ is strictly zero (proline has a fixed $\phi=-65^\circ$ in
the model, which falls outside the $\phi$ interval in our $\beta$-strand
definition). We also note that the two end residues tend to be 
unstructured for all four sequences, with relatively small values of both 
$\ev{\chia(i)}$ and $\ev{\chib(i)}$.

From the single-residue probabilities $\ev{\chia(i)}$ and $\ev{\chib(i)}$, 
one cannot tell whether or not the formation of secondary structure is 
cooperative. To study that for S1, S2 and S4, we calculate the helix-helix 
correlation coefficient for {\em neighboring} residues at $T=299$\,K, as
defined by
\begin{equation}
r^{(\alpha)}_{i\,i+1}=
\frac{C^{(\alpha)}_{i\,i+1}}{\sqrt{C^{(\alpha)}_{ii} C^{(\alpha)}_{i+1\,i+1}}},
\end{equation}
where
\begin{equation}
C^{(\alpha)}_{ij}=\ev{\chia(i)\chia(j)}-\ev{\chia(i)}\ev{\chia(j)}.
\end{equation}
For all three peptides, we find that the largest $r_{i\,i+1}^{(\alpha)}$ 
values occur in the region from $i=4$ to $i=9$ and are in the range 0.3--0.5. 
These values indicate that helix formation is a rather weakly
cooperative process for these peptides. Consequently, the   
free-energy barrier to helix formation should be low, a conclusion
that is in line with the results shown in Fig.~4a.  
For S3, $r_{i\,i+1}^{(\alpha)}$ is about 0.3 or smaller for all $i$.
The analogous strand-strand correlation coefficient $r^{(\beta)}_{i\,i+1}$,
defined in terms of $\chib(i)$, is smaller than 0.25 for all $i$ 
for all the four sequences. 

Another way of analyzing secondary-structure correlations is to  
look at the typical lengths of unbroken $\alpha$-helix and $\beta$-strand
segments. Specifically, we calculate the fraction of conformations,
at fixed $T$, that have at least one unbroken $\alpha$-helix 
($\beta$-strand) stretch with 3 residues or more, which we denote   
by $\la$ ($\lb$). Table~2 shows 
$\la$ and $\lb$ for S1 and S3 at three different temperatures. 
For S1 at $T=299$\,K, we find that $\la=0.12$. This result can be compared 
with what one would expect if the $\chia(i)$ were independent random variables 
with $i$-dependent individual distributions, given by Fig.~5a. In 
this uncorrelated case, it turns out that one would find $\la=0.04$.   
This comparison shows that the correlations are significant but not very
strong. For S3, we find that $\lb=0.04$ at $T=299$\,K. A calculation
analogous to that for S1, shows that $\lb=0.04$ is precisely what 
one would expect in the absence of correlations. Hence, we find that  
secondary-structure correlations are very weak for S3.

Finally, it is also instructive to identify the backbone H bonds 
that are most likely to occur. We consider an H bond formed 
if its energy is $<-\epsilon_\text{hb}^{(1)}/3$.
For S1, we find that the bonds
NH(Asp6)-CO(Asn3) and NH(Asn7)-CO(Asn3) occur in $\approx38$\,\% and
$\approx34$\,\% of the conformations, respectively, at $T=299$\,K, whereas
no other backbone H bond has a frequency of occurrence above 
15\,\%. These results confirm that the $\alpha$-helix seen in low-energy
conformations for S1 is most stable in its N-terminal part. Note also
that in our simulations this helix often starts with a fork-like H 
bonding; the CO(Asn3) group acts as an acceptor for two bonds. For S3, 
there is only one backbone H bond that occurs in more than 15\,\%
of the conformations at $T=299$\,K, namely NH(Asn11)-CO(Ala8) with a 
frequency of occurrence of $\approx21$\,\%. The paucity of H bonds 
underscores the notion that this peptide is highly flexible.   

\paragraph{Two Other Sequences} 
Why do we find a different behavior for S3? A major reason is the different
position of the proline; the proline residue with its special geometry is at 
position 9 in the sequence S3, but at position 4 in S1, S2 and S4. 
To gauge the importance of the proline location, we repeated the  
same calculations for a variant of S3, S$3^\prime$, with Asp4 and 
Pro9 interchanged.  
We find that the behavior of S$3^\prime$ closely 
resembles that of S1, S2 and S4. As an example, we show 
in Fig.~5 the $\alpha$-helix and $\beta$-strand probability
profiles for S$3^\prime$. The S$3^\prime$ profiles are nearly identical 
to those for S1, S2 and S4. In the reshuffling of S1 to 
get S3, the change of proline position thus seems
particularly important.           

We also studied the sequence obtained by interchanging 
Pro4 and Thr9 in S1, which we call S$1^\prime$.
We find that this transposition of S1 leads to a behavior 
similar to that of S3, as is illustrated by Fig.~5, 
which confirms the importance of the position of the proline.  

Neither S$1^\prime$ nor S$3^\prime$ has, to our knowledge,
been studied experimentally.
 
\section*{\centering{Conclusions}}
\label{sec:sum}

We have investigated the solution behavior of four synthetic
peptides, S1--S4, that experimentally have been shown to exhibit
specific adhesion properties to (100) GaAs and Si 
semiconductor substrates. We find that S1, the double mutant S2
and the triple mutant S4 all show a very similar behavior with respect to 
structure as well as thermodynamics. At room temperature, these peptides
are largely unstructured, but have a small but significant
$\alpha$-helix content. For S3, which is a random permutation of 
S1, we find a different behavior. S3 is more flexible than the
other three peptides, with a very small content of both $\beta$-strand and 
$\alpha$-helix structure. The lowest-energy structure we find for S3 is 
not $\alpha$-helical but a $\beta$-hairpin.

In the experiments, S1--S4 all showed good adhesion to GaAs, 
especially S1. The main difference between the peptides was that S3, 
in contrast to the other three, adhered well to Si, too. 

Interestingly, our results suggest a clear difference in 
solution behavior between S3 and the other three peptides. To what extent 
conformational differences can explain the different adhesion properties 
of these peptides remains to be seen. A possible test of this would be
to determine the adhesion properties of the sequence S$3^\prime$, which 
in our model shows a solution behavior similar to that of S1, S2 and S4.   
It would be very interesting to see whether the adhesion properties of 
S$3^\prime$ resemble those of S1, S2 and S4, with similar conformational
properties as S$3^\prime$, or whether they resemble those of S3, with 
83\,\% sequence identity to S$3^\prime$.

\paragraph{Acknowledgments}
This work was in part supported by the German Science Foundation   
(no. JA 483/24-1) and the Swedish Research Council. Support from the
DAAD-STINT Personnel Exchange Programme and 
the John von Neumann Institute for Computing (NIC),
Forschungszentrum J\"ulich (no.\ hlz11) is gratefully acknowledged.  
%

%

\newpage

\parindent 0pt
\parskip 2ex

\begin{center}
{\bf \Large Tables}
\end{center}

\begin{table}[h]
\begin{center}
\caption{The four sequences studied, and their 
PAC values for adsorption to (100) GaAs and 
Si surfaces (from Goede et al.\cite{goede2}).
S2 is a double His$\to$Ala mutant of S1, S3 a random
permutation of S1, and S4 a triple Asn$\to$Ala mutant of S1.}

\vspace{6pt}

\begin{tabular}{llcc}\hline\hline
      &              &\multicolumn{2}{c}{PAC} \\ 
label & sequence     &   GaAs     &   Si      \\ \hline
S1    & AQNPSDNNTHTH &    25\%    &    1\%    \\
S2    & AQNPSDNNTATA &    14\%    &    3\%    \\
S3    & TNHDHSNAPTNQ &    17\%    &   15\%    \\
S4    & AQAPSDAATHTH &    21\%    &    6\%    \\ \hline\hline  
\end{tabular}
\end{center}
\end{table}

\vspace{24pt}

\begin{table}[h]
\begin{center}
\caption{The fraction $\lambda_\alpha$ ($\lambda_\beta$) 
  of conformations
  that have at least one continuous $\alpha$-helix ($\beta$-strand) segment
  of length 3 or more.}

\vspace{6pt}

\begin{tabular}{lrrrrrr}
  \hline\hline & \multicolumn{3}{c}{S1} & \multicolumn{3}{c}{S3} \\
  $T$ (K) & 275 & 299 & 369 & 275 & 299 & 369 \\\hline
  $\lambda_\alpha$ & 0.21 & 0.12 & 0.03 & 0.06 & 0.04 & 0.01 \\
  $\lambda_\beta$ & 0.03 & 0.03 & 0.04 & 0.04 & 0.04 & 0.03 \\
  \hline\hline
\end{tabular}
\end{center}
\end{table}

\newpage

{\Large\bf Figure Captions}

{\bf Fig.~1.}\quad
Temperature dependence of (a) the specific
heat $\Cv=d\ev{E}/dT$ and (b) $d\ev{\Rg}/dT$, for the sequences S1--S4. 
$\Rg$ is the radius of gyration (calculated over all non-H atoms).  

{\bf Fig.~2.}\quad 
Temperature dependence of (a) the $\alpha$-helix 
content $\ev{\na}$ and 
(b) the $\beta$-strand content $\ev{\nb}$, for the sequences S1--S4.
We define a residue as $\alpha$-helical if its Ramachandran 
angles  $\phi$ and $\psi$ satisfy $\phi\in(-90^\circ,-30^\circ)$ and 
$\psi\in(-77^\circ,-17^\circ)$, and $\na$ denotes the fraction of the 10 
inner residues that are $\alpha$-helical.  Similarly, $\nb$ is the fraction of 
the 10 inner residues with Ramachandran angles satisfying
$\phi\in(-150^\circ,-90^\circ)$ and $\psi\in(90^\circ,150^\circ)$.

{\bf Fig.~3.}\quad 
Typical low-energy conformations 
for (a) S1, (b) S2, (c) S3, and (d) S4. These structures 
were obtained as the lowest-energy structures in ten simulated 
annealing runs for each sequence, starting from random conformations. 
In each run, the temperature was decreased geometrically from 
369\,K to 0.7\,K in 100 steps. At each temperature 100\,000 
elementary MC steps were performed.
Drawn with PyMOL.\cite{pymol}

{\bf Fig.~4.}\quad 
Free energies 
$F(\Delta,E)$ calculated as functions of RMSD, $\Delta$,
and energy, $E$, for S1 and S3 at $T=299$\,K. The
reference structure is either an  
$\alpha$-helix or a $\beta$-hairpin (see text). 
The contours are spaced at intervals 
of 1\,$\kB T$. Contours more than 6 $\kB T$ above the minimum 
free energy are not shown. The free energy $F(\Delta,E)$ is defined by 
$P(\Delta,E)\propto\exp(-F(\Delta,E)/\kB T)$, where $P(\Delta,E)$ is 
the joint probability distribution of $\Delta$ and $E$ at temperature $T$. 
(a) RMSD from the $\alpha$-helix for S1 (calculated over residues 5--12). 
(b) RMSD from the $\alpha$-helix for S3 (residues 1--8). 
(c) RMSD from the $\beta$-hairpin for S3 (all residues).
Note that the $x$ scale is different in (c).

{\bf Fig.~5.}\quad 
Secondary-structure profiles for S1--S4, 
S$1^\prime$  (AQNTSDNNPHTH) and 
S$3^\prime$  (TNHPHSNADTNQ) at $T=299$\,K.
(a) The probability that residue $i$ is in the $\alpha$-helix state, 
$\ev{\chia(i)}$, against $i$.   
(b) The probability that residue $i$ is in the $\beta$-strand state, 
$\ev{\chib(i)}$, against $i$. The lines are only guides to the eye.

\newpage

\begin{center}
\epsfig{figure=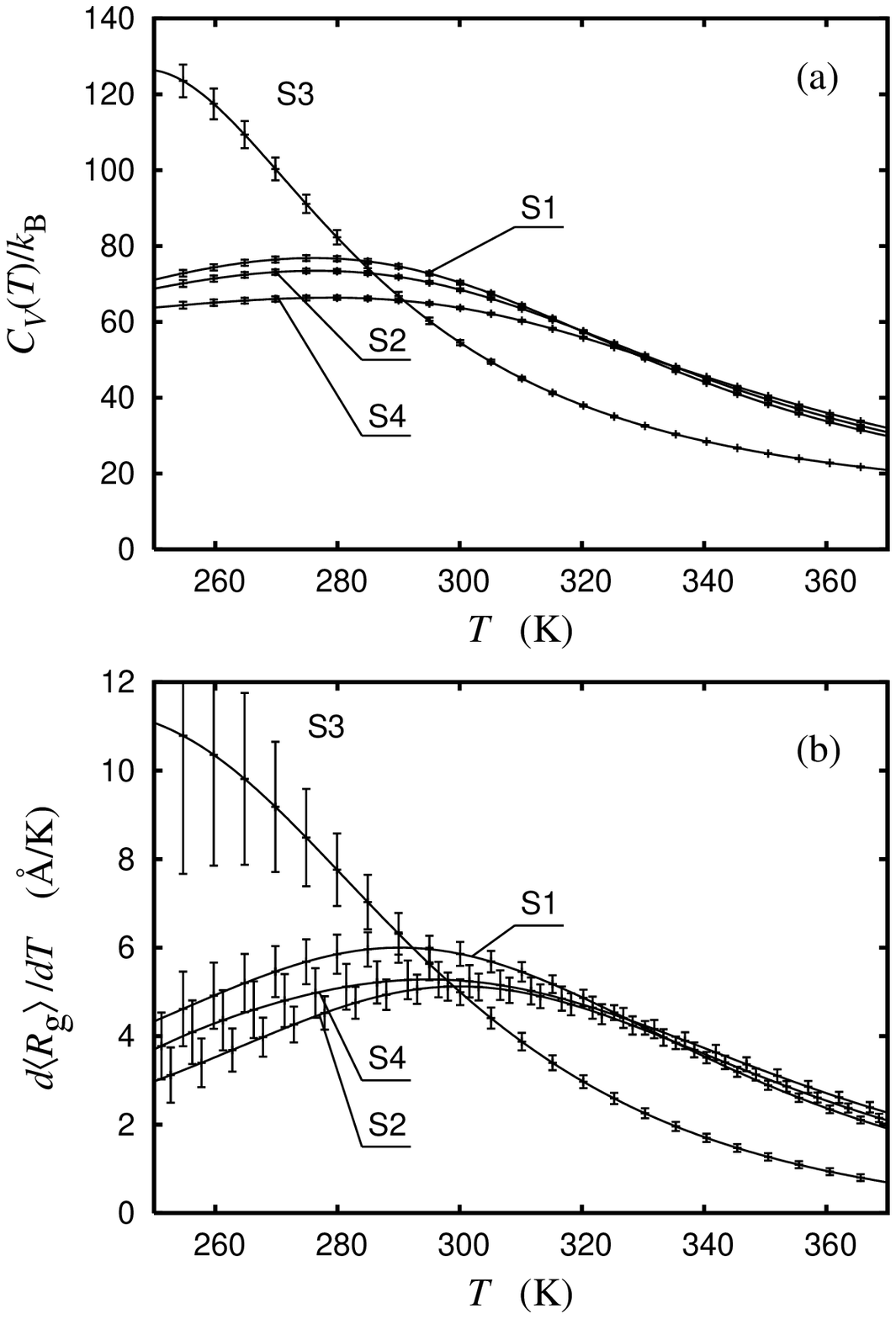,width=8.25cm}
\end{center}

\vspace{-12pt}

\begin{center}
{\bf Fig. 1}
\end{center}

\newpage

\begin{center}
\epsfig{figure=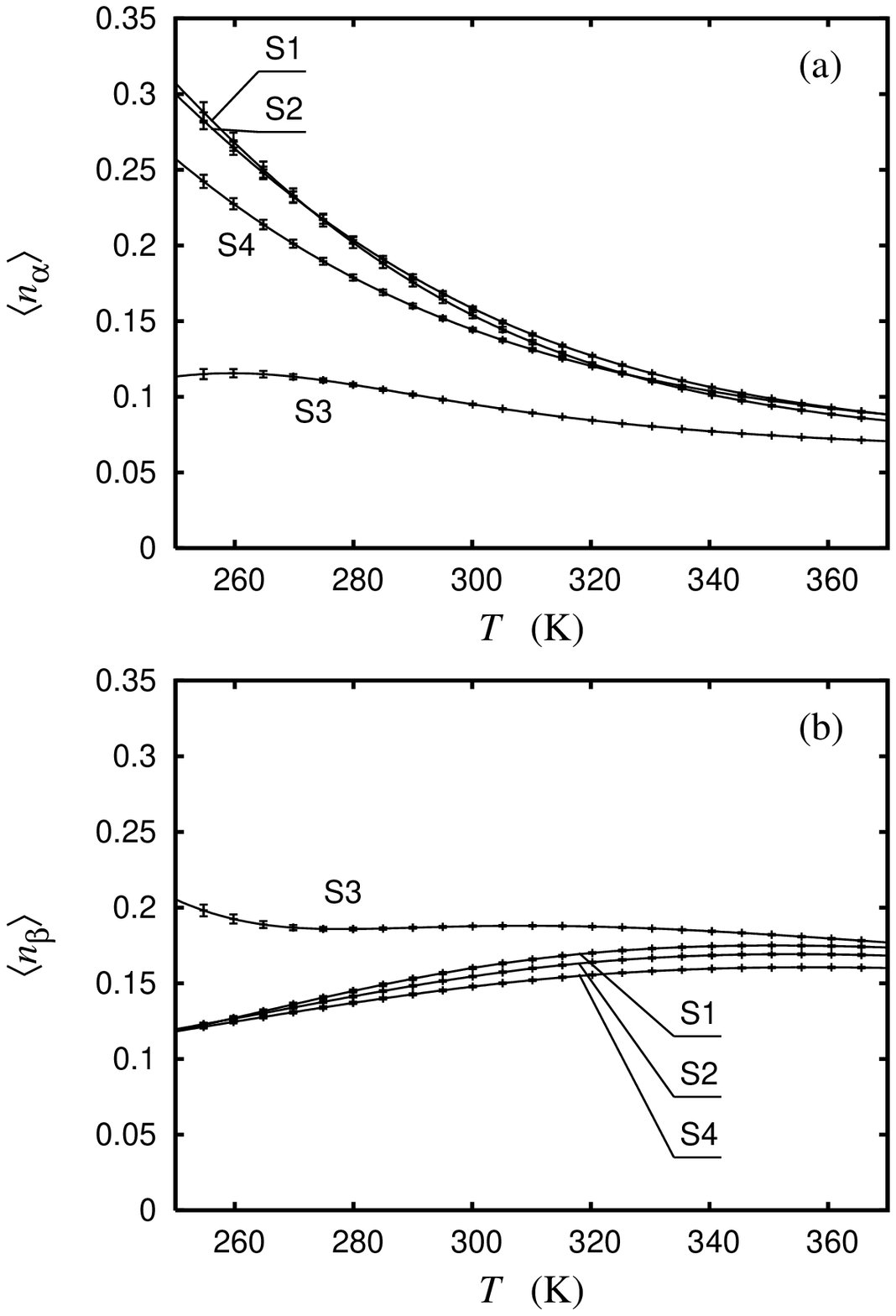,width=8.25cm}
\end{center}

\vspace{-12pt}

\begin{center}
{\bf Fig. 2}
\end{center}

\newpage

\begin{center}
\epsfig{figure=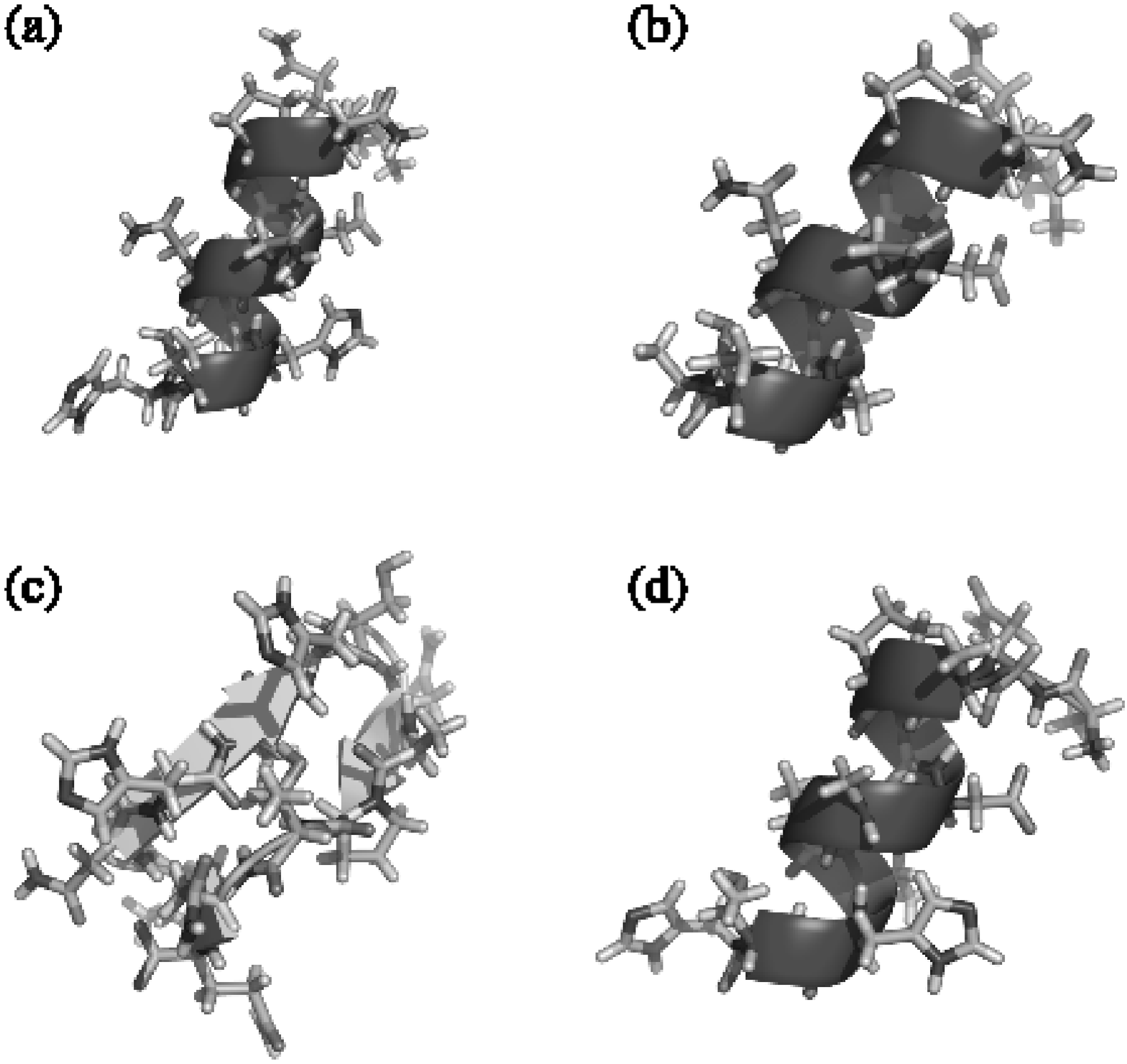,width=8.25cm}
\end{center}

\vspace{-12pt}

\begin{center}
{\bf Fig. 3}
\end{center}

\vspace{48pt}

\begin{center}
\epsfig{figure=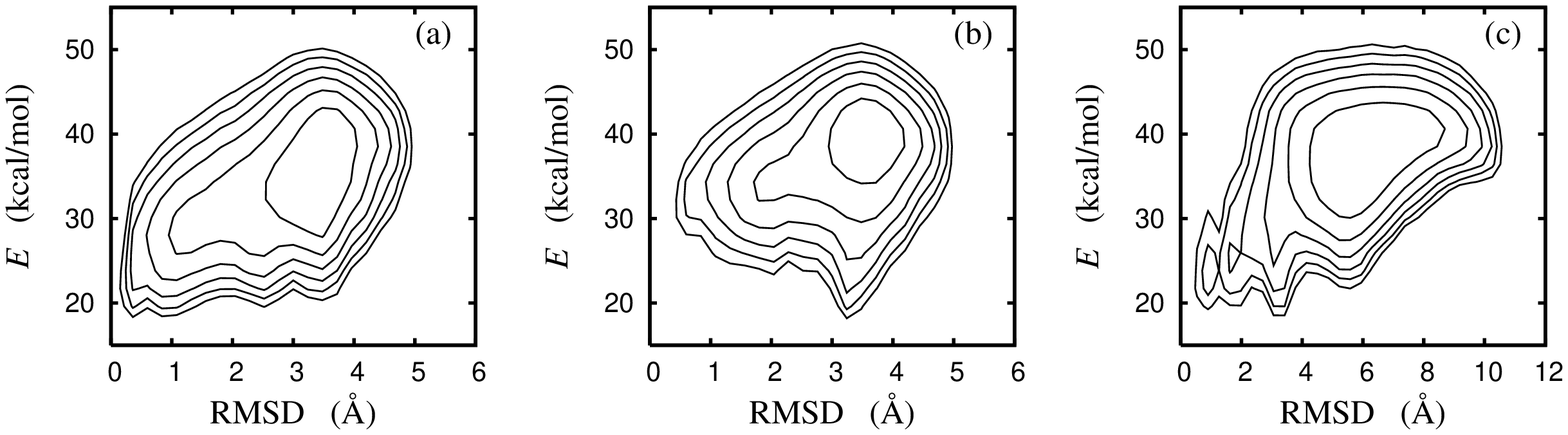,width=16.5cm}
\end{center}

\vspace{-12pt}

\begin{center}
{\bf Fig. 4}
\end{center}

\newpage

\begin{center}
\epsfig{figure=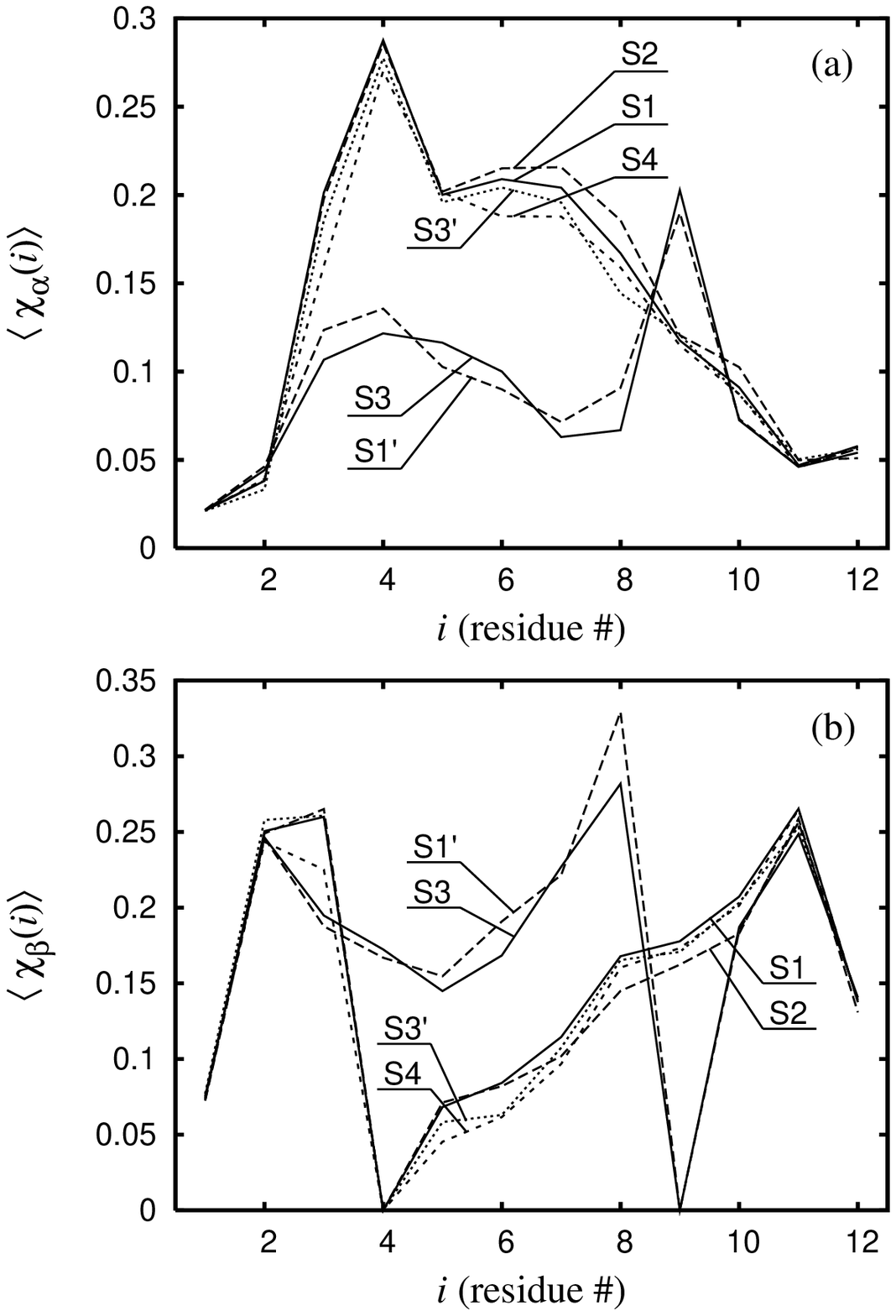,width=8.25cm}
\end{center}

\vspace{-12pt}

\begin{center}
{\bf Fig. 5}
\end{center}

\begin{thebibliography}{199}
%
\bibitem{sarikaya1}
Sarikaya, M.; Tamerler, C.; Jen, A.~K.-Y.; Schulten, K.; Baneyx, F.
{\em Nature Mat.} {\bf 2003}, 2, 577--585.
%
\bibitem{gray}
Gray, J.~J. 
{\em Curr. Opin. Struct. Biol.} {\bf 2004}, 14, 110--115.
%
\bibitem{brown1}
Brown, S.
{\em Nature Biotechnol.} {\bf 1997}, 15, 269--272. 
%
\bibitem{sano}
Sano, K.-I.; Shiba, K.
{\em J. Am. Chem Soc.} {\bf 2003}, 125, 14234--14235.
%
\bibitem{whaley1}
Whaley, S.~R.; English, D.~S.; Hu, E.~L.; Barbara, P.~F.; Belcher, A.~M. 
{\em Nature} {\bf 2000}, 405, 665--668.
%
\bibitem{wang}
Wang, S.; Humphreys E.~S.; Chung S.-Y.; Delduco D.~F.; Lustig, S.~R.;
Wang, H.; Parker, K.~N.; Rizzo, N.~W.; Subramoney, S.; Chiang, Y.-M.;
Jagota, A. 
{\em Nature Mat.} {\bf 2003}, 2, 196--200. 
%
\bibitem{goede1}
Goede, K.; Busch, P.; Grundmann, M. 
{\em Nano Lett.} {\bf 2004}, 4, 2115--2120.
%
\bibitem{goede2}
Goede, K.; Grundmann, M.; Holland-Nell, K.; Beck-Sickinger, A.~G.
{\em Langmuir} {\bf 2006}, 22, 8104--8108.
%
\bibitem{dyson}
Dyson, H.~J.; Wright, P.~E. 
{\em Curr. Opin. Struct. Biol.} {\bf 2002}, 12, 54--60.
%
\bibitem{shoemaker}
Shoemaker, B.~A.; Portman, J.~J.; Wolynes, P.~G.
{\em PNAS} {\bf 2000}, 97, 8868--8873.
%
\bibitem{irbaeck1}
Gupta, N.; Irb\"ack, A. 
{\em J.\ Chem.\ Phys.} {\bf 2004}, 120, 3983--3989.
%
\bibitem{willett1}
Willett, R.~L.; Baldwin, K.~W.; West, K.~W.; Pfeiffer, L.~N. 
{\em PNAS} {\bf 2005}, 102, 7817--7822.
%
\bibitem{peelle1}
Peelle, B.~R.; Krauland, E.~M.; Wittrup, K.~D.; Belcher, A.~M.
{\em Langmuir} {\bf 2005}, 21, 6929--6933.
%
\bibitem{irbaeck2}
Irb\"ack, A.; Samuelsson, B.; Sjunnesson, F.; Wallin, S. 
{\em Biophys.\ J.} {\bf 2003}, 85, 1466--1473.
%
\bibitem{irbaeck4}
Irb\"ack, A.; Mohanty, S. 
{\em Biophys.\ J.} {\bf 2005}, 88, 1560--1569.
%
\bibitem{gbcj1}
G\"{o}ko\u{g}lu, G.; Bachmann, M.; \c{C}elik, T.; Janke, W. 
{\em Phys.\ Rev.\ E} {\bf 2006}, 74, 041802.
%
\bibitem{ecepp}
N\'emethy, G.; Gibson, K.~D.; Palmer K.~A.; Yoon, C.~N.; Paterlini, G.;
Zagari, A.; Rumsey, S.; Scheraga, H.~A. 
{\em J. Phys. Chem.} {\bf 1992}, 96, 6472--6484.
%
\bibitem{cormack}
Cormack, A.~N.; Lewis, R.~J.; Goldstein, A.~H.
{\em J. Phys. Chem. B} {\bf 2004}, 108, 20408--20418.
%
\bibitem{liang}
Liang, T.; Walsh, T.~R.
{\em Phys. Chem. Chem. Phys.} {\bf 2006}, 8, 4410--4419.
%
\bibitem{schulten1}
Braun, R.; Sarikaya, M.; Schulten, K. 
{\em J.\ Biomater.\ Sci., Polym.\ Ed.} {\bf 2002}, 13, 747--757.
%
\bibitem{vrbova1}
Vrbov{\`a}, T.; Whittington, S.~G. 
{\em J.\ Phys.\ A} {\bf 1996}, 29, 6253--6264.
%
\bibitem{bj1}
Bachmann, M.; Janke, W.
{\em Phys.\ Rev.\ Lett.} {\bf 2005}, 95, 058102.
%
\bibitem{bj2}
Bachmann, M.; Janke, W.
{\em Phys.\ Rev.\ E} {\bf 2006}, 73, 041802.
%
\bibitem{bj3}
Bachmann, M.; Janke, W. 
{\em Phys.\ Rev.\ E} {\bf 2006}, 73, 020901(R).
%
\bibitem{muthukumar} 
Muthukumar, M. {\em J. Chem. Phys.} {\bf 1995}, 103, 4723--4731.
%
\bibitem{bratko}
Bratko, D.; Chakraborty, A.~K.; Shakhnovich, E.~I.
{\em Chem. Phys. Lett.} {\bf 1997}, 280, 46--52. 
%
\bibitem{golumbfskie}
Golumbfskie, A.~J.; Pande, V.~S.; Chakraborty, A.~K.
{\em PNAS} {\bf 1999}, 96, 11707--11712. 
%
\bibitem{bogner1}
Bogner, T.; Degenhard, A.; Schmid, F.
{\em Phys.\ Rev.\ Lett.} {\bf 2004}, 93, 268108.
%
\bibitem{kriksin}
Kriksin, Y.~A.; Khalatur, P.~G.; Khokhlov, A.~R.
{\em J. Chem . Phys.} {\bf 2005}, 122, 114703.
%
\bibitem{moghaddam}
Moghaddam, M.~S.; Chan, H.~S. 
{\em J. Chem . Phys.} {\bf 2006}, 125, 164909.
%
\bibitem{zhdanov}
Zhdanov, V.~P.; Kasemo, B. {\em Proteins} {\bf 1998}, 30, 168--176. 
%
\bibitem{castells}
Castells, V.; Yang, S.; Van Tassel, P.~R. 
{\em Phys. Rev. E} {\bf 2002}, 65, 031912. 
%
\bibitem{Marinari:92}
Marinari, E.; Parisi, G. 
\textit{Europhys. Lett.} \textbf{1992}, 19, 451--458.
%
\bibitem{Lyubartsev:92}
Lyubartsev, A.~P.; Martsinovski, A.~A.; Shevkunov, S.~V.;
Vorontsov-Velyaminov, P.\ N.  
\textit{J. Chem. Phys.} \textbf{1992}, 96, 1776--1783.
%
\bibitem{BGS}
Favrin, G.; Irb\"ack, A.; Sjunnesson, F. 
\textit{J.\ Chem.\ Phys.} \textbf{2001}, 114, 8154--8158.
%
\bibitem{irbaeck5}
Irb\"ack, A.; Mohanty, S.\ {\em J.\ Comput.\ Chem.} {\bf 2006}, 27, 1548--1555.
The PROFASI package is freely available at 
http://cbbp.thep.lu.se/activities/profasi.
%
\bibitem{ferrenberg}
Ferrenberg, A.~M.; Swendsen, R.~H.
\textit{Phys. Rev. Lett.} \textbf{1989}, 63, 1195--1198.
%
\bibitem{miller}
Miller, R.~G. \textit{Biometrika} \textbf{1974}, 61, 1--15. 
%
\bibitem{kirkpatrick}
Kirkpatrick, S.; Gelatt Jr., C.~D.; Vecchi, M.~P.
\textit{Science} \textbf{1983}, 220, 1297--1301.
%
\bibitem{pymol}
DeLano, W.~L. The PyMOL Molecular Graphics System (2002) on World Wide
Web http://www.pymol.org.
%
\end{thebibliography}
\end{document}